\documentclass[preprint,12pt,authoryear]{elsarticle}

\usepackage[margin=1in]{geometry}
\usepackage[T1]{fontenc}
\usepackage[utf8]{inputenc}
\usepackage{lmodern}
\usepackage{booktabs}
\usepackage{tabularx}
\usepackage{longtable}
\usepackage{array}
\usepackage{ragged2e}
\newcolumntype{Y}{>{\RaggedRight\hspace{0pt}\arraybackslash}X}
\newcolumntype{P}[1]{>{\RaggedRight\hspace{0pt}\arraybackslash}p{#1}}
\usepackage{enumitem}
\usepackage{csquotes}
\usepackage{xcolor}
\usepackage{xurl}
\usepackage[hidelinks,pdfencoding=auto,psdextra]{hyperref}
\usepackage{orcidlink}

\hypersetup{%
  pdftitle={From Code-Centric to Intent-Centric Software Engineering: A Reflexive Thematic Analysis of Generative AI, Agentic Systems, and Engineering Accountability},
  pdfauthor={Elyson De La Cruz},
  pdfkeywords={generative artificial intelligence; agentic systems; software engineering; artificial intelligence pair programming; reflexive thematic analysis; software engineering agents}
}

\journal{Journal of Systems and Software}

\begin{document}

\begin{frontmatter}

\title{From Code-Centric to Intent-Centric Software Engineering: A Reflexive Thematic Analysis of Generative AI, Agentic Systems, and Engineering Accountability}

\author[inst1]{Elyson De La Cruz\texorpdfstring{\,\orcidlink{0009-0006-5599-506X}}{}}
\affiliation[inst1]{organization={University of the Cumberlands}, country={United States}}
% Corresponding author email should be entered in the journal submission system.

\begin{abstract}
Generative artificial intelligence (GenAI) and agentic systems are moving software engineering from code-centric production toward intent-centric human-agent work in which natural language, repository context, tools, tests, and governance shape delivery. Prior studies examine code generation, AI pair programming, and software engineering agents, but less is known about how public technical discourse and peer-reviewed evidence together frame the profession's near-term transition. This study addresses that gap through a reflexive thematic analysis (RTA) dominant and interpretative phenomenological analysis (IPA) informed public-discourse and document analysis. The corpus combines peer-reviewed software engineering and AI literature, technical benchmarks, public talks and interviews, essays, product-facing technical announcements, and X-originated discourse from prominent AI and software engineering voices. Sources were organized through a corpus register, codebook, coding matrix, theme-to-source traceability table, DOI/reference audit, and reproducibility protocol. The analysis shows that GenAI lowers the cost of producing plausible code while increasing the importance of intent specification, context curation, architecture knowledge, verification, security, provenance, governance, and accountable human judgment. The findings indicate that software engineering is becoming less about isolated code authorship and more about supervising, validating, and governing socio-technical systems of humans, agents, tools, and evidence gates. This matters because speed-focused adoption can accumulate hidden technical debt and accountability gaps, whereas bounded autonomy can preserve quality, security, maintainability, and trust.
\end{abstract}

\begin{keyword}
generative AI \sep agentic systems \sep software engineering \sep AI pair programming \sep reflexive thematic analysis \sep interpretative phenomenological analysis \sep software engineering agents \sep human-AI collaboration \sep software architecture \sep software quality
\end{keyword}

\end{frontmatter}

\section{Introduction}
Software engineering is entering a new phase in which the act of writing code is no longer the exclusive center of software work. Large language models (LLMs) trained on source code demonstrated that natural language prompts can produce executable programs, with Codex and related systems showing early capability on function-level programming tasks \citep{vaswani2017attention,brown2020language,chen2021codex,li2022competition,lyu2025automatic}. Subsequent work moved the evaluation unit from toy problems toward repository-level issue resolution, as reflected in SWE-bench and related benchmarks that test whether language models can resolve real GitHub issues in real codebases \citep{jimenez2023swebench,xia2025demystifying}. At the same time, empirical and practitioner-facing studies show that AI pair-programming systems create new usability, quality, security, and trust challenges \citep{peng2023impact,zhou2025problems,eshraghian2024emotional,pearce2022asleep,tihanyi2024secure}.

This transformation has revived a long-standing software engineering question: what is essential, and what is accidental, in software development? Brooks argued that no single technology removes the essential difficulty of software engineering because the core challenges reside in complexity, conformity, changeability, and invisibility \citep{brooks1987nosilverbullet}. GenAI reduces some accidental friction, especially boilerplate generation, translation between languages, test scaffolding, documentation drafting, and local debugging. However, it does not remove the essential burden of understanding stakeholder intent, architectural tradeoffs, security consequences, data constraints, organizational accountability, or long-term maintainability. Technical-debt research reinforces this concern by showing that short-term implementation compromises create future maintenance and evolution costs that require explicit management \citep{li2015technicaldebt,alves2016technicaldebt}.

The near-term evolution of software engineering is therefore unlikely to produce a simple substitution of developers by agents. A more plausible trajectory is a reallocation of engineering effort. Software engineers will spend less time producing first-draft code and more time specifying intent, curating context, supervising agents, designing evidence gates, validating generated artifacts, managing architectural knowledge, and assuring compliance. The unit of engineering will shift from a human editing a file to a human-agent system operating across requirements, code, tests, build pipelines, vulnerability scanners, deployment workflows, and operational telemetry.

This paper develops an RTA-dominant and IPA-informed analysis of this shift. The study is not a participant interview study; rather, it reports a completed public-discourse and document analysis that can be audited through the accompanying corpus register, coding matrix, and source-to-theme traceability files. Reflexive thematic analysis is appropriate because the present corpus includes heterogeneous public artifacts and requires interpretive pattern development rather than claims of coding reliability or measurement objectivity \citep{braun2006thematic,braun2019reflecting,braun2020canuse,braunclarke2024rtarg}. IPA remains useful as a sensitizing lens because software engineers and AI researchers are interpreting changing professional identity, responsibility boundaries, authorship, and control \citep{smithosborn2004ipa,smith2021ipa,vanscoy2015ipa}. The paper therefore uses IPA idiographically, researcher by researcher and role by role, before using RTA to develop cross-case themes.

The central argument is that GenAI and agentic systems are producing an \emph{intent-centric} software engineering paradigm. In this paradigm, humans increasingly express desired outcomes, constraints, quality attributes, policies, and acceptance evidence, while agents generate, test, repair, and explain candidate implementations. The trajectory is not autonomy without humans. It is bounded autonomy inside engineered control loops.

\section{Research Questions}
The study is organized around three research questions:

\begin{description}[leftmargin=1.2cm,style=nextline]
    \item[RQ1:] How are GenAI and agentic systems shifting the locus of software engineering work from code production to intent specification, verification, and governance?
    \item[RQ2:] What dichotomies structure practitioners' interpretation of this shift, particularly between augmentation and automation, speed and quality, autonomy and accountability, and democratization and expertise?
    \item[RQ3:] What maturity stages best characterize the near-term evolution of software engineering roles, practices, methods, and research priorities under increasing agentic capability?
\end{description}

\section{Contributions}
This article makes four contributions. First, it provides a methodologically bounded qualitative account of how prominent AI researchers, software engineering researchers, and practitioner commentators frame the transition from code-centric to intent-centric software work. Second, it contributes a reproducible public-discourse corpus protocol that distinguishes peer-reviewed evidence, technical preprints, and public thought-leadership artifacts. Third, it develops a dichotomy-centered interpretation of the transition: GenAI lowers the marginal cost of plausible code while increasing the importance of verification, architecture knowledge, governance, and accountability. Fourth, it offers a maturity-stage pathway and research agenda for empirical studies of AI-augmented software engineering in organizations.

\section{Research Design Workflow}
The study used the following workflow to align the research questions, corpus, analytic method, trustworthiness procedures, and supplemental evidence package. The workflow is reported textually to preserve journal compatibility and to make each analytic decision auditable:

\begin{enumerate}[leftmargin=1.2cm]
    \item \textbf{Define the phenomenon and analytic stance.} The phenomenon is the shift from code-centric software authorship to intent-centric human-agent software engineering. The epistemological stance is interpretivist and constructivist: public discourse is treated as situated meaning-making, not as neutral measurement.
    \item \textbf{Specify corpus boundaries.} The corpus includes peer-reviewed publications, public talks, YouTube interviews, public essays, product-facing technical announcements, and X-originated posts from prominent AI researchers and software engineering practitioners whose public claims address GenAI, coding agents, software architecture, reasoning, safety, or developer work.
    \item \textbf{Apply inclusion and exclusion criteria.} Included artifacts must be attributable, dateable, publicly accessible or archived, and substantively relevant to software engineering or agentic AI\@. Excluded artifacts include unattributed reposts, anonymous summaries, purely promotional claims without technical content, and claims that cannot be traced to a public source or publication.
    \item \textbf{Capture and normalize evidence.} Each artifact is logged with author, role, organization at time of source, date, source type, URL or DOI, transcript status, key excerpts, and relevance to the research questions. YouTube content is analyzed from transcripts when available; X posts are archived through direct links or verifiable secondary reproductions when direct access is unstable.
    \item \textbf{Conduct first-cycle semantic coding.} Initial codes stay close to the artifact language, such as ``natural language as programming,'' ``agents as workflow actors,'' ``architecture as context,'' ``verification burden,'' ``reasoning limits,'' and ``governance risk.''
    \item \textbf{Conduct second-cycle latent coding.} Codes are interpreted at a higher level to identify the meanings, assumptions, tensions, and professional identities implied by each artifact.
    \item \textbf{Develop idiographic profiles.} Each prominent researcher or practitioner is first analyzed as an individual case. This step preserves the IPA-informed attention to situated perspective before cross-case abstraction.
    \item \textbf{Generate cross-case RTA themes.} Themes are developed recursively by comparing semantic codes, latent meanings, publications, and discrepant cases across researchers, organizations, and source types.
    \item \textbf{Triangulate against peer-reviewed literature.} Public discourse themes are checked against empirical software engineering, code-generation, software architecture, AI pair-programming, security, and qualitative-methods literature.
    \item \textbf{Audit and refine claims.} The manuscript and supplemental corpus were reviewed for methodological congruence, unsupported forecasting, overstatement of agent autonomy, missing counter-evidence, journal build quality, and citation integrity.
\end{enumerate}

\begin{table}[htbp]
\centering
\caption{Research design workflow and quality controls}
\label{tab:workflow}
\begin{tabularx}{\textwidth}{P{0.22\textwidth}Y Y}
\toprule
\textbf{Workflow phase} & \textbf{Research action} & \textbf{Quality control} \\
\midrule
Problem formulation & Define the software engineering paradigm shift and research questions. & Align claims with RQs and avoid deterministic forecasting. \\
Corpus construction & Select public discourse artifacts and peer-reviewed publications. & Use inclusion/exclusion criteria, source log, date, author, URL/DOI, and archive status. \\
Case-level analysis & Analyze each thought leader as an idiographic case. & Preserve context, role, organization, and source type before cross-case abstraction. \\
RTA theme development & Generate semantic codes, latent codes, candidate themes, and final themes. & Maintain reflexive memos and revise themes recursively rather than seeking mechanical inter-rater reliability. \\
Triangulation & Compare public claims with software engineering and AI research. & Distinguish peer-reviewed evidence from public thought leadership. \\
Trustworthiness & Establish credibility, dependability, confirmability, and transferability. & Use audit trail, discrepant-case analysis, thick description, member-reflection option, and peer debriefing. \\
Claim audit & Convert analysis into propositions, matrix, maturity pathway, and research agenda. & Check that each interpretive claim is supported by a source trail and by an appropriate evidence class. \\
\bottomrule
\end{tabularx}
\end{table}

\section{Background and Thought-Leadership Framing}

\subsection{From software as crafted source code to software as learned behavior}
Karpathy's ``Software 2.0'' framing captured an important conceptual shift: a growing portion of system behavior is no longer hand-coded in procedural logic but learned from data and represented in model weights \citep{karpathy2017software2}. This idea does not replace software engineering; it expands the artifact boundary. Modern software increasingly includes source code, model weights, prompts, retrieval pipelines, evaluation harnesses, policy constraints, and telemetry feedback loops. As a result, the software artifact has become a hybrid of deterministic code and probabilistic behavior.

This hybridization now extends to the software development process itself. LLMs are not only embedded in products; they are embedded in the engineering workflow. The developer uses models to generate code, interpret logs, write tests, refactor modules, produce documentation, and reason about defects. The development environment therefore becomes a socio-technical system in which human intention, model output, repository context, tool execution, and organizational policy interact.

\subsection{The enduring value of architecture knowledge}
Parnas's information-hiding principle remains relevant because modular decomposition is still a primary defense against uncontrolled complexity \citep{parnas1972criteria}. AI-generated code does not automatically respect hidden design decisions, implicit constraints, or long-lived architecture rationale. Ozkaya argues that architecture knowledge is central to the responsible use of GenAI in software development because architecture decisions encode why a system is structured in a particular way, not merely what code exists \citep{ozkaya2023architecture}. This observation is central to the present argument: agentic coding without architecture-aware context can accelerate implementation while weakening design coherence.

\subsection{From coding assistants to software engineering agents}
The first wave of AI coding assistance emphasized code completion, function generation, documentation, and single-turn interaction. Empirical studies show benefits, but also issues around misunderstandings, incorrect suggestions, over-reliance, and workflow friction \citep{peng2023impact,zhou2025problems,eshraghian2024emotional}. The second wave emphasizes agents that can plan, search a repository, edit multiple files, run tests, repair failures, and iterate. Recent research on the future of software development with GenAI similarly frames the field as moving from isolated tool support toward broader process and workflow transformation \citep{sauvola2024future}. SWE-bench made this transition visible by testing whether language models could resolve real repository issues rather than only synthesize standalone functions \citep{jimenez2023swebench,xia2025demystifying}. Recent software engineering agent research further shows that agent design involves tool access, planning, memory, environment feedback, repository context, and evaluation strategy \citep{xia2025demystifying,fan2023llmse,zhang2026survey,lyu2025automatic}.

This transition changes the engineering question. The first question was: can a model generate useful code? The next question is: can a human-agent system produce accountable, secure, maintainable, evolvable software under real constraints?

\section{Methodology: RTA-Dominant, IPA-Informed Public-Discourse Analysis}

\subsection{Methodological position}
The methodology uses reflexive thematic analysis as the primary analytic method and IPA as a secondary sensitizing lens. This choice is methodologically important. A pure IPA study would normally require first-person participant accounts generated through interviews or diaries and would focus on how participants make sense of lived experience \citep{smithosborn2004ipa,smith2021ipa}. The current manuscript, however, analyzes public discourse artifacts and publications from prominent AI researchers and practitioners. RTA is therefore more defensible because it supports theoretically flexible, interpretive, pattern-based analysis across heterogeneous qualitative materials \citep{braun2006thematic,braun2019reflecting,braun2020canuse}.

The study is positioned within an interpretivist and constructivist paradigm. It does not assume that public posts or interviews transparently reveal stable attitudes. Instead, it treats them as situated discourse produced by researchers, executives, practitioners, and tool builders who occupy different positions in the AI ecosystem. The purpose is not to count how many people agree with a claim. The purpose is to interpret how influential actors frame the software engineering transition, where their framings converge or diverge, and how these framings align with empirical software engineering evidence.

\subsection{Study type and unit of analysis}
The study design is a qualitative public-discourse and document analysis with an RTA analytic strategy. Document analysis is appropriate when public texts, reports, interviews, and records are treated as analyzable qualitative evidence rather than as background context alone \citep{bowen2009document}. The unit of analysis is the \emph{meaning-bearing artifact}, which may be a talk segment, interview answer, X post, public essay, technical announcement, or peer-reviewed publication passage. The secondary unit of analysis is the \emph{thought-leader case}, defined as the body of traceable artifacts associated with a prominent researcher or practitioner.

\subsection{Citation alignment and evidence hierarchy}
The reference strategy follows a layered evidence hierarchy. Peer-reviewed journal articles and top-tier archival conference papers are used to support scholarly claims about software engineering, technical debt, security, code review, and qualitative methodology. Technical preprints are retained only when they define influential benchmarks, model artifacts, or rapidly emerging systems for which a peer-reviewed version is not yet available. Public talks, X-originated posts, interviews, and essays are treated as public-discourse data for thematic interpretation, not as empirical validation of technical performance. This separation is important because the study analyzes public thought leadership while grounding its scholarly claims in peer-reviewed software engineering, AI, security, and qualitative-methods literature.

\subsection{Corpus construction}
The corpus was curated in three layers. Layer 1 contains peer-reviewed research evidence, including studies on LLM code generation, software engineering agents, AI pair programming, security of AI-generated code, architecture knowledge, code review, technical debt, and qualitative methodology. Layer 2 contains technical preprints and benchmark artifacts retained because the field is moving faster than journal publication cycles. Layer 3 contains public thought-leadership artifacts, including public talks, YouTube interviews, podcasts, essays, product-facing technical posts, and X-originated posts. The supplemental corpus package records metadata, source dates, source locators, evidence roles, archive status, provisional codes, theme links, researcher case profiles, and peer-review alignment notes. Long copyrighted transcripts and full X post text are not redistributed; the corpus uses source locators and paraphrased meaning-unit rationales.

\begin{table}[htbp]
\centering
\small
\caption{Corpus construction protocol}
\label{tab:corpus}
\begin{tabularx}{\textwidth}{P{0.24\textwidth}Y Y}
\toprule
\textbf{Corpus layer} & \textbf{Included evidence} & \textbf{Defensive use in the manuscript} \\
\midrule
Peer-reviewed literature & Journal articles, conference papers, journal articles or conference papers with identifiable authorship, DOI or proceedings identifiers, and relevance to software engineering or AI agents. & Used to support empirical and technical claims. \\
Public thought leadership & YouTube talks, interviews, podcasts, public essays, product technical posts, and direct or archived X posts. & Used as discourse artifacts and sensitizing concepts, not as proof of empirical outcomes. \\
Traceability records & Source log, date, author, affiliation at time of artifact, URL/DOI, transcript excerpt, code label, memo, and theme linkage. & Used to demonstrate analytic transparency and allow readers to follow theme development. \\
Disconfirming sources & Skeptical talks, limitations-focused publications, safety critiques, failure cases, and governance incidents. & Used to avoid hype bias and preserve analytic balance. \\
\bottomrule
\end{tabularx}
\end{table}

The priority public cases include Andrej Karpathy, Andrew Ng, Ipek Ozkaya, Simon Willison, Thomas Dohmke, Boris Cherny, Demis Hassabis, Jeff Dean, Koray Kavukcuoglu, David Silver, John Jumper, Pushmeet Kohli, Yann LeCun, Francois Chollet, Fei-Fei Li, and xAI-associated researchers such as Igor Babuschkin, Yuhuai Wu, Jimmy Ba, Christian Szegedy, Greg Yang, Zihang Dai, and Guodong Zhang. These names are not treated as a popularity list. They are selected because their public positions represent different analytic poles: software-as-natural-language, agentic workflow, architecture governance, practitioner operation, scaling and infrastructure, reasoning and reinforcement learning, safety, and principled skepticism about current LLM limitations.

\subsection{Supplemental corpus and reproducibility package}
The supplemental corpus is submitted as a companion package rather than embedded in the article body. It contains a corpus register, codebook, coding matrix, theme table, researcher case profiles, exclusion log, data dictionary, DOI/reference audit, methodology alignment audit, and SHA-256 manifest. The package supports three forms of transparency. First, readers can trace each theme to specific source identifiers. Second, readers can distinguish peer-reviewed evidence from public discourse artifacts. Third, future researchers can replicate or extend the corpus by adding participant interviews, repository traces, agent logs, or longitudinal organizational cases. This design aligns with the expectation that claims be supported by evidence and that research data be made available or explained when unavailable.

\subsection{Sampling strategy}
The study uses purposive theoretical sampling. Public artifacts are selected because they illuminate the phenomenon, not because they represent a statistically generalizable population. Maximum variation is sought across organizational position, research tradition, and discourse stance. The sample intentionally includes frontier-lab researchers, software engineering researchers, AI-native tool builders, AI skeptics, and practitioner analysts. This sampling design supports theoretical breadth while preserving enough idiographic attention to avoid collapsing all researchers into a single ``AI thought leader'' category.

The final sample is bounded to public and citable artifacts. It intentionally does not include private interviews, proprietary repositories, confidential agent logs, or non-public organizational data. This boundary increases reproducibility because all included source artifacts can be traced by readers, but it also limits the study's ability to represent ordinary practitioners whose experiences are not visible in public discourse.

\subsection{Analytic procedure}
The analysis follows Braun and Clarke's recursive RTA logic while adding an IPA-informed case interpretation step \citep{braun2006thematic,braun2019reflecting,braunclarke2024rtarg}. The procedure is not a mechanical reliability exercise. RTA treats researcher subjectivity as a resource to be managed reflexively rather than eliminated through coding consensus alone.

\begin{enumerate}[leftmargin=1.2cm]
    \item \textbf{Familiarization.} Read or view artifacts repeatedly; annotate transcript passages, claims, metaphors, cautionary statements, and implied role assumptions.
    \item \textbf{Semantic coding.} Code explicit statements, such as ``natural language programming,'' ``agents run tests,'' ``architecture knowledge matters,'' or ``LLMs lack planning.''
    \item \textbf{Latent coding.} Interpret underlying meanings, such as changing authorship, shifting accountability, automation anxiety, governance debt, or scale optimism.
    \item \textbf{Idiographic case memoing.} Write a short profile for each thought leader, preserving role, institutional context, technical tradition, and public stance.
    \item \textbf{Theme construction.} Cluster codes into candidate themes and compare them against the research questions.
    \item \textbf{Theme review.} Test themes against the corpus, peer-reviewed literature, and disconfirming cases. Revise themes that are too broad, too vendor-specific, or insufficiently supported.
    \item \textbf{Theme definition and naming.} Define each theme as an interpretive claim, not a topic label. For example, ``verification becomes the central engineering act'' is stronger than ``testing.''
    \item \textbf{Manuscript integration.} Map each theme to the dichotomy matrix, maturity-stage pathway, and research agenda.
\end{enumerate}

\subsection{Coding framework}
The coding framework used for first-cycle and second-cycle analysis is shown in Table~\ref{tab:codebook}. The final theme table in the supplement records how these codes were consolidated into cross-case interpretations.

\begin{table}[htbp]
\centering
\small
\caption{Initial RTA and IPA-informed coding framework}
\label{tab:codebook}
\begin{tabularx}{\textwidth}{P{0.22\textwidth}Y Y}
\toprule
\textbf{Code family} & \textbf{Indicative codes} & \textbf{Interpretive focus} \\
\midrule
Programming interface & Natural language programming, Software 3.0, prompt as specification, vibe coding. & How actors reframe what it means to program. \\
Agentic workflow & Reflection, planning, tool use, multi-agent coordination, terminal/browser/editor access. & How single-turn prompting becomes iterative workflow engineering. \\
Architecture and context & Architecture knowledge, design rationale, system constraints, repository context. & How local code generation interacts with system-level coherence. \\
Verification and evidence & Tests, review burden, security scanning, provenance, pull-request evidence, benchmarks. & How generated artifacts become acceptable or unacceptable. \\
Professional identity & Authorship, expertise, deskilling, supervision, seniority, junior learning. & How practitioners reinterpret their role and value. \\
Governance and risk & Permissions, sandboxing, audit logs, prompt injection, policy-as-code, rollback. & How organizations bound agent autonomy. \\
Capability skepticism & Reasoning limits, world models, abstraction, planning, memory, generalization. & How skeptical researchers constrain hype-based forecasts. \\
\bottomrule
\end{tabularx}
\end{table}

\subsection{Trustworthiness and quality criteria}
Trustworthiness is addressed through credibility, dependability, confirmability, and transferability, consistent with qualitative rigor principles and RTA reporting guidance \citep{lincoln1985naturalistic,nowell2017trustworthiness,braunclarke2024rtarg}. Credibility is supported by triangulating public discourse with peer-reviewed software engineering evidence and by actively searching for disconfirming cases. Dependability is supported by a documented source log, coding memos, theme-development record, and revision history. Confirmability is supported by separating source excerpts from analytic interpretation and by maintaining reflexive memos about the researcher's assumptions. Transferability is supported by thick description of artifact type, researcher role, organizational context, and source date.

The study does not use inter-rater reliability as a primary quality criterion because that would conflict with reflexive TA's interpretive assumptions. Instead, a structured adversarial audit was used for theme challenge, evidence-role checking, and claim discipline. The audit focused on whether interpretations were traceable to the corpus and appropriately bounded by evidence class.

\subsection{Ethical and legal considerations}
The public corpus contains public materials, but public availability does not remove ethical responsibility. The study quotes public posts sparingly, cites sources transparently, and avoids implying private intent beyond what the artifact supports. For X posts, the source log distinguishes direct posts, archived screenshots, and secondary reports. The corpus excludes proprietary code, credentials, secrets, customer data, restricted operational information, and private agent transcripts. Public X-originated artifacts are treated cautiously because posts can be deleted, edited, decontextualized, or redistributed through secondary reporting.

\subsection{Methodological limitations}
The design has three important limitations. First, public thought leadership is performative and strategic; it may reflect persuasion, product positioning, or institutional incentives as much as personal belief. Second, public discourse overrepresents prominent voices and underrepresents ordinary engineers, maintainers, QA specialists, and security reviewers. Third, the maturity pathway is an analytic projection rather than a predictive model; stage timing will vary by organizational context, regulatory pressure, toolchain maturity, and risk tolerance. These limitations are mitigated by distinguishing evidence types, preserving discrepant cases, and framing findings as propositions for empirical testing.

\subsection{Reference validation and evidence hierarchy}
To meet journal-submission expectations, the study separates three evidentiary classes. First, peer-reviewed journal and conference publications support scholarly claims about software engineering, technical debt, code generation, software architecture, AI pair programming, AI-generated code security, and qualitative methodology. Priority was given to indexed and highly selective venues, including \emph{Journal of Systems and Software}, \emph{Empirical Software Engineering}, \emph{ACM Transactions on Software Engineering and Methodology}, \emph{Science}, \emph{Communications of the ACM}, \emph{Automated Software Engineering}, \emph{Information Technology \& People}, \emph{Qualitative Research in Psychology}, and \emph{International Journal of Qualitative Methods}. Second, arXiv records and benchmark papers are used only when they are foundational to the technical discourse and when no more stable peer-reviewed substitute yet exists, such as Codex, SWE-bench, and public productivity evidence. Third, public talks, interviews, essays, X-originated posts, and product discourse are treated as qualitative data artifacts rather than as peer-reviewed evidence.

DOI-bearing references were checked for DOI syntax, metadata consistency, and source suitability; core DOI metadata was also cross-checked against bibliographic metadata services where available. References that could not be supported by stable metadata or that duplicated stronger peer-reviewed evidence were removed or demoted to public-corpus status. This hierarchy is important because the study analyzes public thought leadership, but its scholarly claims must remain anchored in peer-reviewed software engineering and qualitative research literature.

\section{The Core Dichotomy}
The dominant tension is not simply human versus AI\@. It is acceleration versus accountability. GenAI accelerates the creation of plausible artifacts. Software engineering remains responsible for deciding whether those artifacts satisfy requirements, preserve architecture, respect security constraints, and improve system value.

\begin{table}[htbp]
\centering
\caption{Dichotomies shaping GenAI and agentic software engineering}
\label{tab:dichotomy}
\begin{tabularx}{\textwidth}{P{0.23\textwidth}Y Y}
\toprule
\textbf{Dichotomy} & \textbf{Acceleration pole} & \textbf{Accountability pole} \\
\midrule
Augmentation vs automation & AI completes, explains, refactors, and scaffolds code. & Humans remain accountable for intent, acceptance, risk, and consequences. \\
Speed vs quality & Agents increase artifact throughput and reduce blank-page effort. & Generated artifacts require verification, maintainability review, and security assessment. \\
Prompting vs architecture & Natural language becomes a primary interface for software work. & Architecture knowledge, constraints, and design rationale must shape prompts and agent context. \\
Local change vs system evolution & Agents can solve local tasks and repository issues. & Systems evolve through dependencies, policies, data contracts, operations, and long-term maintainability. \\
Democratization vs expertise & More users can create working prototypes and automate routine tasks. & Expert review becomes more valuable because non-experts may not detect subtle defects. \\
Autonomy vs governance & Agents can plan, act, run tools, and iterate. & Organizations need identity, permissions, logs, test gates, approval workflows, and rollback controls. \\
\bottomrule
\end{tabularx}
\end{table}

The dichotomy is productive rather than binary. The future of software engineering will be shaped by teams that exploit the acceleration pole while engineering disciplined controls around the accountability pole.

\section{Findings and Interpretive Themes}
The following eight themes are analytic interpretations developed from the coded public-discourse corpus and triangulated against peer-reviewed software engineering literature. The supplemental coding matrix links each theme to source identifiers, provisional codes, analytic memos, and disconfirming notes. The themes should therefore be read as transparent qualitative propositions rather than as statistically generalizable findings.

Table~\ref{tab:theme_support} summarizes the source support for each theme. Support counts do not imply statistical weight; they indicate traceability breadth across the curated corpus. The subsection structure below mirrors the table exactly so that each tabulated theme is explained in the narrative findings.

\begin{table}[htbp]
\centering
\small
\caption{Theme support and evidence-role controls}
\label{tab:theme_support}
\begin{tabularx}{\textwidth}{P{0.10\textwidth}Y P{0.17\textwidth}Y}
\toprule
\textbf{Theme} & \textbf{Short label} & \textbf{Source support} & \textbf{Evidence control} \\
\midrule
T1 & Natural language as programming interface & 7 artifacts & Interpreted as public-discourse framing and triangulated with code-generation literature. \\
T2 & Agentic workflow over single prompt & 16 artifacts & Anchored in software-engineering-agent and benchmark literature where available. \\
T3 & Human as architect, verifier, and accountable operator & 21 artifacts & Supported by architecture, code review, technical debt, and AI pair-programming evidence. \\
T4 & Democratization versus professionalization & 11 artifacts & Treated as discourse interpretation, not as a labor-market performance claim. \\
T5 & Reasoning, planning, and autonomy remain contested & 14 artifacts & Balanced with skeptical and capability-oriented sources to avoid deterministic forecasting. \\
T6 & Safety, governance, and provenance debt & 20 artifacts & Triangulated with security, qualitative auditability, and governance-related sources. \\
T7 & AI systems engineering and infrastructure convergence & 25 artifacts & Interpreted as a systems trajectory rather than a claim that full autonomy is achieved. \\
T8 & Essential versus accidental complexity and technical debt & 9 artifacts & Anchored in Brooks, Parnas, technical debt, code review, and maintainability evidence. \\
\bottomrule
\end{tabularx}
\end{table}

\subsection{Theme 1: Natural language becomes a programming interface}
The first theme captures a shift in the interface through which software work is initiated. Natural language increasingly functions as a practical control surface for code drafting, explanation, refactoring, test generation, and documentation. This does not make programming identical to ordinary conversation. Rather, it changes the first act of programming from writing syntactically precise source code to expressing intent, constraints, examples, and acceptance conditions in a form that an AI system can use.

The theme is grounded in public discourse around software as a new kind of natural-language-mediated activity and triangulated with technical work showing that transformer-based systems can generate code from natural language prompts \citep{vaswani2017attention,brown2020language,chen2021codex,li2022competition,lyu2025automatic}. The practical implication is that intent clarity becomes more scarce as code production becomes less scarce. Requirements engineering, example-driven specifications, property-based tests, and domain-specific constraints therefore become more valuable, not less valuable.

\subsection{Theme 2: Agentic workflow replaces the single prompt as the unit of work}
The second theme shifts attention from isolated prompting to agentic workflows. Early code-generation studies demonstrated that models could synthesize functions or solve bounded programming tasks. Current public product artifacts and software engineering agent research move the unit of analysis toward a loop: plan, retrieve context, edit files, run commands, observe test results, repair errors, and prepare a pull request or other reviewable artifact.

This workflow view is consistent with repository-level benchmarks and emerging software engineering agent studies that evaluate performance against real issues, real codebases, and tool-mediated execution environments \citep{jimenez2023swebench,xia2025demystifying,github2026copilotagent,openai2025codex,anthropic2026claudecode,google2025antigravity}. The engineering implication is that organizations should not ask only whether a model can write code. They should ask whether the human-agent workflow produces auditable evidence that the change is correct, safe, maintainable, and aligned with repository conventions.

\subsection{Theme 3: Humans become architects, verifiers, and accountable operators}
The third theme directly addresses the changing locus of expertise. As agents produce more implementation artifacts, human value shifts toward architecture reasoning, context selection, verification, acceptance decisions, and accountability. Human engineers are not removed from the system; they move into the role of accountable operators who supervise model behavior, interpret ambiguous requirements, judge tradeoffs, and decide whether generated artifacts should enter the software supply chain.

This theme is consistent with classic and contemporary software engineering evidence. Parnas's modularity principle and Ozkaya's architecture-knowledge argument both show that code is only one visible layer of deeper design rationale and system structure \citep{parnas1972criteria,ozkaya2023architecture}. Code review and technical-debt studies further show that long-term quality depends on human judgment, socio-technical coordination, and future maintenance consequences \citep{bacchelli2013code,li2015technicaldebt,alves2016technicaldebt}. The result is a role shift: senior engineering judgment becomes more central as code generation becomes more automated.

\subsection{Theme 4: Democratization and professionalization advance together}
The fourth theme explains a key tension in the corpus. GenAI democratizes software creation by lowering the barrier to prototypes, scripts, internal tools, and first-draft implementations. At the same time, professional software engineering becomes more demanding because production systems require testing, maintainability, security, provenance, architectural coherence, and accountability. In this sense, the same tools that make building easier also increase the importance of disciplined engineering practice.

Empirical and practitioner-facing studies of AI pair programming support this dual movement. Developers can experience productivity gains, reduced blank-page effort, and affective benefits, but they also encounter incorrect suggestions, usability friction, misplaced trust, and uncertainty about responsibility \citep{peng2023impact,zhou2025problems,eshraghian2024emotional}. The finding should not be read as a labor-market prediction. It is a professional-practice interpretation: non-experts can build more, but expert review and governance become more important when more people and agents can produce plausible software artifacts.

\subsection{Theme 5: Reasoning, planning, and autonomy remain contested}
The fifth theme introduces a necessary counterweight to adoption enthusiasm. Public and technical discourse often frames coding agents as increasingly capable planners. However, the corpus also contains cautionary positions that distinguish task success from robust reasoning, long-horizon planning, world understanding, and reliable generalization. The study therefore treats agent autonomy as emergent, bounded, and uneven across contexts rather than as a settled capability.

This theme matters because software engineering often operates under incomplete requirements, legacy constraints, security consequences, and organizational ambiguity. Current benchmarks and agent studies are valuable, but they still simplify parts of the socio-technical environment in which professional software evolves \citep{jimenez2023swebench,xia2025demystifying,fan2023llmse,zhang2026survey}. The practical implication is that maturity should be evaluated by task class. Routine, well-specified changes may support higher automation, while novel design, ambiguous requirements, safety-critical systems, and cross-organizational tradeoffs require stronger human supervision.

\subsection{Theme 6: Safety, governance, and provenance debt become first-order concerns}
The sixth theme identifies a new form of debt created by agentic development: governance and provenance debt. As agents gain access to repositories, terminals, browsers, cloud sandboxes, package managers, test runners, and pull-request workflows, organizations must decide who or what is authorized to act, what data agents may see, how outputs are logged, and what evidence is required before changes merge.

Security research shows that AI-generated code can contain vulnerabilities, and that model outputs vary with task, prompt, language, and evaluation method \citep{pearce2022asleep,tihanyi2024secure}. Public product artifacts also show that coding agents increasingly operate across multiple tools and execution environments \citep{github2026copilotagent,openai2025codex,anthropic2026claudecode,google2025antigravity}. The implication is that governance cannot be an afterthought. Agentic workflows require identities, permissions, sandboxing, review gates, audit logs, provenance records, rollback procedures, and policy-as-code controls.

\subsection{Theme 7: Software engineering converges with AI systems engineering and infrastructure}
The seventh theme broadens the unit of analysis from code to infrastructure. Agentic software engineering is not merely a better autocomplete system. It connects models, prompts, repositories, vector or retrieval systems, tool APIs, CI/CD pipelines, security scanners, telemetry, policy engines, and deployment environments. Software engineering therefore converges with AI systems engineering, where the artifact includes code, model behavior, orchestration logic, evaluation harnesses, and operating constraints.

This convergence is visible in work on GenAI-driven software development and software engineering agents, which frames future practice around integrated workflows rather than isolated code suggestions \citep{sauvola2024future,fan2023llmse,zhang2026survey,lyu2025automatic}. The implication is organizational as much as technical. Mature teams will need shared practices for model selection, context engineering, evaluation, infrastructure cost management, data governance, incident response, and continuous monitoring of human-agent delivery pipelines.

\subsection{Theme 8: Essential complexity and technical debt do not disappear}
The eighth theme anchors the study in a long-running software engineering tradition. GenAI can reduce accidental complexity by automating boilerplate, translation, scaffolding, documentation, and some debugging work. However, it does not remove essential complexity: understanding user needs, negotiating tradeoffs, designing abstractions, managing change, preserving modularity, securing systems, and maintaining software over time.

This theme is grounded in Brooks's distinction between essential and accidental difficulty and in Parnas's argument that modular decomposition affects understandability and changeability \citep{brooks1987nosilverbullet,parnas1972criteria}. It is also reinforced by technical-debt and code-review research, which shows that short-term implementation choices can create future costs that are difficult to detect at the time of change \citep{bacchelli2013code,li2015technicaldebt,alves2016technicaldebt}. The implication is that organizations should not treat generated code as free capacity. They should treat it as a faster path to candidate artifacts that still require architectural stewardship, quality gates, and long-term ownership.

Together, the eight themes support the central proposition of the manuscript: GenAI makes creation cheaper, while agentic software engineering makes accountability more important. The following maturity pathway translates that proposition into a staged view of organizational capability without assigning rigid calendar years to each stage.

\section{Maturity Pathway for Agentic Software Engineering}

Rather than treating the next five years as a deterministic calendar forecast, this section presents a maturity pathway for agentic software engineering. The stages are ordered by capability, governance maturity, and organizational adoption patterns. The horizon labels are approximate and should not be read as predictions that a given capability will mature in a specific year. Tool capability, regulation, organizational risk tolerance, repository readiness, and enterprise integration are evolving too quickly for a precise calendar forecast. Agentic coding is already operational in mainstream developer environments: public product artifacts in the corpus describe coding agents that can be assigned issues or pull-request work, operate in cloud or sandboxed environments, read codebases, edit files, run commands, execute tests, and use editor, terminal, and browser surfaces \citep{github2026copilotagent,openai2025codex,anthropic2026claudecode,google2025antigravity}. The near-term research problem is therefore not whether agentic systems will enter software engineering. The more important question is how organizations mature from individual agent-assisted delivery toward governed, evidence-producing, architecture-aware, and bounded-autonomous software engineering.

\begingroup\footnotesize
\setlength{\tabcolsep}{3pt}
\begin{longtable}{@{}P{0.13\textwidth}P{0.13\textwidth}P{0.22\textwidth}P{0.25\textwidth}P{0.18\textwidth}@{}}
\caption{Maturity pathway from agent-assisted coding to governed agentic software engineering}\label{tab:trajectory}\\
\toprule
\textbf{Maturity stage} & \textbf{Approximate horizon} & \textbf{Dominant capability} & \textbf{Engineering shift} & \textbf{Primary risk} \\
\midrule
\endfirsthead
\toprule
\textbf{Maturity stage} & \textbf{Approximate horizon} & \textbf{Dominant capability} & \textbf{Engineering shift} & \textbf{Primary risk} \\
\midrule
\endhead
Stage 1: Individual agent assistance & Current to near term & Developers use copilots and coding agents in IDEs, terminals, cloud sandboxes, and pull-request workflows. & AI supports code drafting, debugging, documentation, testing, codebase navigation, and first-pass issue resolution. & Over-trust, weak review, hallucinated dependencies, shallow tests, and unclear ownership of generated changes. \\
Stage 2: Team-level governed agent use & Near term & Teams formalize agent permissions, repository instructions, coding standards, evidence gates, and review rules. & Agent use shifts from individual productivity to governed team workflow with repeatable operating norms. & Shadow-agent use, inconsistent policy, architecture drift, audit gaps, and uneven skill development. \\
Stage 3: Lifecycle-integrated agent workflows & Mid term & Specialized agents support testing, security, documentation, release, modernization, and compliance tasks. & Software delivery becomes a coordinated human-agent workflow across the lifecycle. & Conflicting agents, brittle orchestration, hidden technical debt, and accountability ambiguity. \\
Stage 4: Intent-centric engineering platforms & Mid to longer term & Requirements, architecture, policy, code, tests, deployment, and telemetry become connected inputs to agentic workflows. & Requirements and architecture decisions become machine-readable context for agents; acceptance evidence becomes continuous and traceable. & Misalignment between stakeholder intent and automated implementation, especially in regulated or safety-critical systems. \\
Stage 5: Bounded-autonomous software engineering & Longer term & Routine and well-specified change classes can be handled by supervised autonomous engineering workflows. & Humans focus on architecture, risk, product judgment, socio-technical coordination, governance, and novel system design. & Accountability ambiguity, junior developer deskilling, governance debt, and concentration of expertise. \\
\bottomrule
\end{longtable}
\endgroup

The pathway is expected to unfold unevenly across organizations. Digitally mature organizations may already exhibit Stage 2 or Stage 3 practices, while regulated, legacy, or low-governance environments may remain at Stage 1 for longer. The model is therefore intended as a diagnostic and research framing tool, not a universal maturity timetable.

\section{Implications for Practice}

\subsection{For software engineering leaders}
Leaders should avoid measuring AI adoption only by lines of generated code or short-term velocity. Better indicators include cycle time, escaped defects, review burden, security findings, test coverage quality, rollback frequency, maintainability, developer learning, and stakeholder value. Leaders should define acceptable AI use, generated-code review standards, data-handling constraints, and escalation criteria for agent autonomy.

\subsection{For architects}
Architects should make architecture knowledge computable and accessible. Agentic systems need decision records, dependency maps, interface contracts, threat models, non-functional requirements, and diagrams that can be retrieved and reasoned over. Architecture documentation should not be static decoration; it should be operational context for human-agent engineering.

\subsection{For security and compliance teams}
Security teams should assume that AI-generated code can be useful and unsafe at the same time. Security controls should include static analysis, dependency scanning, secret detection, software bill of materials generation, adversarial prompt testing, provenance logging, and policy-as-code controls. Agent actions should have identities, permissions, logs, and approval requirements.

\subsection{For educators}
Software engineering education should not respond by banning GenAI\@. It should teach students how to work with AI while preserving fundamentals. Curricula should emphasize code reading, debugging, testing, architecture, secure coding, requirements reasoning, model limitations, and professional accountability. Students should learn to ask: What did the tool produce, why might it be wrong, how can I verify it, and what are the consequences if it fails?

\section{Research Agenda}
The field needs stronger evidence on long-term and system-level effects. Short-term productivity studies are valuable, but the next research frontier is maintainability, security, socio-technical adaptation, and organizational governance. Priority research questions include:

\begin{enumerate}[leftmargin=1.2cm]
    \item How does AI-generated code affect technical debt after six, twelve, and twenty-four months?
    \item What review practices best detect subtle AI-generated defects?
    \item Which repository structures make agentic development safer and more effective?
    \item How do junior developers learn when AI generates first drafts?
    \item What governance models balance agent autonomy with human accountability?
    \item How should software engineering benchmarks account for architecture, security, maintainability, and organizational context?
\end{enumerate}

\section{Threats to Validity and Limitations}
This manuscript reports a qualitative public-discourse and document analysis, not a participant-based IPA interview study. The methodology is therefore RTA-dominant and IPA-informed. The themes synthesize current literature and curated public thought-leadership artifacts rather than elicited interview transcripts. The findings should be treated as transparent qualitative propositions that can be tested and refined through future interview, diary, repository, and organizational case-study research.

The public corpus introduces source-related limitations. YouTube interviews and X posts may be partial, performative, edited, promotional, or unstable over time. X posts in particular require careful archiving because access, deletion, and platform-level changes can affect traceability. The study mitigates this concern by requiring a source log, artifact metadata, archive status, and clear distinction between peer-reviewed evidence and public discourse.

The rapid pace of model and tooling development also limits forecast certainty. A capability that is experimental in 2026 may become routine by 2028, while legal, security, labor, or regulatory pressures may slow adoption. Another limitation is tool heterogeneity. ``GenAI'' and ``agentic systems'' cover diverse systems with different capabilities, costs, privacy properties, and integration patterns. The analysis therefore focuses on software engineering dynamics rather than any single vendor or model. Finally, productivity evidence remains mixed across tasks and contexts. A tool that accelerates one developer on a greenfield task may slow another developer working in a complex legacy repository.

\section{Conclusion}
The next phase of software engineering will not be defined by the disappearance of engineers. They will be defined by the reconfiguration of engineering work around intent, context, evidence, and accountability. GenAI reduces the cost of producing plausible code, while agentic systems expand the scope of automated action across repositories and toolchains. This makes software engineering more, not less, dependent on architecture, verification, security, governance, and human judgment.

The core dichotomy is acceleration versus accountability. Organizations that embrace acceleration without accountability will accumulate hidden risk. Organizations that reject acceleration will lose learning opportunities and productivity potential. The strategic path is to engineer bounded autonomy: agentic workflows that are powerful enough to reduce routine work, but constrained enough to preserve human responsibility, system coherence, and public trust.

\section*{CRediT authorship contribution statement}
Elyson De La Cruz: Conceptualization, methodology, investigation, data curation, formal analysis, writing -- original draft, writing -- review and editing, project administration.

\section*{Data availability}
The supplemental corpus package is prepared for DOI-backed archival release through Zenodo and transparent version control through GitHub. The repository-ready package includes the public-discourse corpus register, coding matrix, codebook, theme table, maturity pathway table, researcher case profiles, exclusion log, data dictionary, DOI/reference audit, methodology alignment audit, reproducibility protocol, and SHA-256 manifest. No private participant data, proprietary source code, credentials, customer data, or non-public organizational materials were collected. Full YouTube transcripts and complete X post text are not redistributed because of copyright, platform instability, and ethical traceability concerns; the supplement provides source locators, source metadata, evidence-role labels, and paraphrased meaning-unit rationales.

\section*{Ethics statement}
The study uses publicly accessible discourse artifacts and published literature. It does not involve private human-subject interviews, intervention, confidential repositories, or non-public organizational records. Any future extension involving interviews, repository walkthroughs, or agent logs should receive institutional ethics review and should exclude credentials, secrets, customer data, proprietary source code, and sensitive operational telemetry.

\section*{Declaration of generative AI and AI-assisted technologies}
Generative AI was used as a consistency and writing support tool during manuscript preparation. The author remains responsible for all claims, citations, interpretations, source validation, and final content.

\section*{Declaration of competing interest}
The author declares no competing interests. This statement should be updated before submission if any author has employment, consulting, investment, advisory, product, or funding relationships with AI coding-assistant vendors, cloud providers, software engineering tool vendors, or organizations discussed in the public corpus.

\section*{Funding}
This research did not receive any specific grant from funding agencies in the public, commercial, or not-for-profit sectors.

\appendix

\section{Supplemental materials submitted separately}
The supplemental package contains machine-readable CSV and JSONL files and a human-readable workbook. The primary files are: corpus register, coding matrix, codebook, theme table, researcher case profiles, exclusion log, data dictionary, DOI/reference audit, methodology alignment audit, reproducibility protocol, and SHA-256 manifest. These files are designed to support reviewer audit, source-level traceability, and future replication.

\bibliographystyle{elsarticle-harv}
\bibliography{refs}

\end{document}